\documentclass{optica-article}

\journal{opticajournal} 

\articletype{Research Article}

\usepackage{lineno}
\usepackage{subcaption}
\usepackage{float}
\usepackage{layouts}
\usepackage{siunitx}
\usepackage{amsmath}
\usepackage{physics}
\usepackage{leftindex}

\newcommand{\Rabisig}[0]{\Omega_\text{mm}}
\newcommand{\Rabiest}[0]{\hat{\Omega}_\text{mm}}
\newcommand{\Rb}[0]{\leftindex^{87} {\text{Rb}}}

\begin{document}

\title{Background-free calibrated electric-field imaging with Rydberg-state fluorescence and Autler-Townes splitting}

\author{Gabriel Ko,\authormark{1,2*}, Wiktor Krokosz,\authormark{1,2} Mateusz Mazelanik,\authormark{2} Wojciech Wasilewski,\authormark{1,2} and Michał Parniak\authormark{1,2}}

\address{\authormark{1} Faculty of Physics, University of Warsaw, L. Pasteura 5, 02-093 Warsaw, Poland\\
\authormark{2} Centre for Quantum Optical Technologies, Centre of New Technologies, University of Warsaw, S. Banacha 2c, 02-097 Warsaw, Poland\\}

\email{\authormark{*}g.ko@cent.uw.edu.pl} 




\begin{abstract*} 
We demonstrate a spatially resolved method for imaging millimeter-wave (mmWave) electric fields using Rydberg-state fluorescence in a warm atomic vapor. By utilizing a multi-photon ladder excitation scheme, we leverage a specific decay channel that remains dark in the absence of the mmWave field, resulting in high-contrast imaging with effectively zero background. Absolute calibration of the local electric field is achieved by reconstructing the Autler–Townes splitting of the Rydberg resonance across the imaging volume. To ensure robust field extraction across a wide dynamic range--including regimes where spectral features are not fully resolved--we employ a steady-state analysis based on the Gorini–Kossakowski–Sudarshan–Lindblad (GKSL) master equation. We apply this technique to visualize standing-wave interference patterns within a vapor cell and demonstrate the ability to engineer local field distributions using structured dielectric reflectors. This approach provides a versatile and self-calibrating platform for the diagnostic imaging of high-frequency electromagnetic fields and the characterization of mmWave-optical interfaces.
\end{abstract*}
\section{Introduction}

Rydberg atoms couple strongly and selectively to microwave and terahertz radiation, enabling precision electromagnetic‑field sensing, communications, and related metrology. In warm vapor cells, techniques based on electromagnetically induced transparency (EIT) \cite{boller_observation_1991,lukin_quantum_1999,fleischhauer_electromagnetically_2005, sedlacek_microwave_2012, 6910267, anderson_continuous-frequency_2017, stelmashenko_measuring_2020} and related dark‑resonance phenomena have been widely deployed to extract field frequency, phase, and amplitude, and to realize atomic receivers.\cite{anderson_Rydberg_2020,fancher_Rydberg_2021,borowka_sensitivity_2022,PhysRevApplied.23.044037,adams_Rydberg_2020,manchaiah,saffman_quantum_2010,mohapatra_coherent_2007, Raithel_polarization}

In recent decades, efforts have been made to capture the spatial information typically lost in probe-laser monitoring techniques. As fluorescence occurs locally, spatial information of the field present at the Rydberg atoms can be recovered. To that end, work has been done in the capturing of Rydberg atoms' fluorescence to image microwave and millimeter-wave (mmWave) fields with sub-wavelength resolution. \cite{fan_subwavelength_2014, wade_real-time_2017, downes_full-field_2020,schlossberger_two_dimensional_2025,angle-of-arrival,prajapati_investigation_2024,pati_millimeter_2026,chen_terahertz_2022,li_dual-cameras_2024,gurtler_imaging_2003, li_room_2024, holloway_sub-wavelength_2014, Chen:26} 

In this work, we specifically choose to use fluorescence, not coinciding with any laser wavelengths used in the experiment. This results in excellent SNR, since fluorescence at our chosen wavelength can only occur when the mmWave field drives the otherwise forbidden transition. This is in contrast to probe-fluorescence-based imaging, in which measurements are made by monitoring the difference in fluorescence intensity as the probe is scanned through resonance.

We extend fluorescence‑based imaging techniques based on localized Autler-Townes (AT) splitting of a Rydberg resonance using a normally dark transition. Specifically, we first take images of the fluorescence while scanning one of the lasers around the Rydberg dark state when in in the presence of mmWave.  Then, we visualize the AT splitting by constructing a composite image of the fluorescence intensity with respect to the laser detuning. This splitting then allows us to extract the local Rabi frequency along the laser and provides a calibrated, position‑resolved measurement of the mmWave field as well as relative Rydberg population within the optical beams inside the vapor cell.


\section{Experimental Setup}



\begin{figure}[h]
    \centering
    \includegraphics[width=0.8\textwidth]{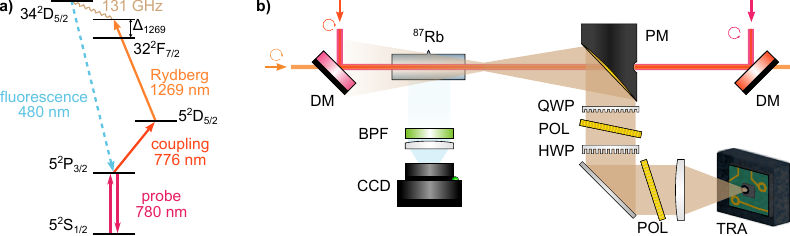}
  \caption{\textbf{a)} Level diagram in $\leftindex^{87}{\text{Rb}}$ relevant to the experiment. \textbf{b)} Snapshot of key components in the setup. Probe laser and 131 GHz beam enter from the right, while the coupling and Rydberg lasers enter from the left. Dichroic mirrors (DM) are placed to align the lasers and prevent laser light from entering the laser emission setup. All fields are circularly polarized to address transitions of the highest dipole moment. The mmWave transceiver (TRA) emits our mmWave and passes through a collimator, polarizers (POL), Half Wave Plate (HWP), and Quarter Wave Plate (QWP) for polarization control. It is then focused into the cell via a Parabolic Mirror (PM). The CCD captures 480 nm fluorescence through a Band Pass Filter (BPF) and camera objective.}
    \label{fig:levels-setup}
\end{figure}




\subsection{Setup Description}

The atomic states and equipment used are visualized and condensed in Fig. \ref{fig:levels-setup}. The 780 nm probe laser is locked using a beat-note lock with a frequency-doubled kHz-linewidth laser source (NKT Photonics, 1560 nm). The coupling and Rydberg lasers, 776 nm and 1269 nm, respectively, both use a cavity transfer lock using the same 1560 nm laser. 
The selection of locking points goes as follows. The lasers are aligned to be co-linear through the use of dichroic mirrors for the 780 nm and 776 nm lasers and a standard IR mirror for the 1269 nm laser. A tilted band-pass filter (BPF), not shown in Fig. \ref{fig:levels-setup}, at 780 nm $\pm10$ nm is introduced in the 776 nm beam path to weakly pick off the probe onto a photodetector for monitoring of the probe absorption via probe laser scan. The probe laser is scanned to monitor the absorption spectrum of the D2 line, while the coupling laser frequency is tuned to resonance, where the EIT between the two lasers is strongest via alignment and polarization adjustments. The Rydberg laser is then brought to resonance, with alignment and polarization adjusted similarly. The probe and coupling laser frequencies are locked where the EIT from their respective two-photon excitation was strongest, then the Rydberg laser is locked where the electromagnetically induced absorption (EIA) is strongest. All lasers are focused to a $\SI{700}{\um}$ beam diameter with Rayleigh length \textasciitilde10 cm for the red lasers and \textasciitilde7 cm for the 1269 nm laser. This Rayleigh length was selected to ensure uniformity of the beams within the 50 mm length of the glass cell.

The mmWave source comes from an industrial transceiver, TRA\textunderscore120\textunderscore045, at 130.72 GHz. This is then weakly collimated using a lens made of High Impact Polystyrene (HIPS), which has an approximate refractive index of n=1.5 for mmWaves. This mmWave beam is then focused into the setup via a parabolic mirror (PM) with a through-hole (MPD229H-M01). The fluorescence is observed in our 50 mm $\Rb$ cell through an Edmund optics 486 nm BPF (\#65-085) and camera objective on a CCD camera (ac4096-11gm). 


\subsection{Technique}

The experiment is done as follows in Fig.\ref{fig:levels-setup}\textbf{a}. From the ground state in $\Rb$, $5^2S_{1/2}$, a 3 photon ladder excitation scheme through states $5^2P_{3/2}$, $5^2D_{5/2}$, and $32^2F_{7/2}$ is used to prepare the Rydberg atoms via the probe (780.24 nm), coupling (776.98 nm), and Rydberg (1269.03 nm) lasers respectively. The mmWave at 131 GHz is propagated co-linearly into the cell with the lasers to excite the energy of the Rydberg atoms to state $34^2D_{5/2}$, where the strongest fluorescence path falls to $5^2P_{3/2}$ via 480 nm emission. 

These aforementioned states have a corresponding energy sourced from the Alkali.ne Rydberg Calculator (ARC) \cite{SIBALIC2017319}: $5^2S_{1/2} = 33691.1 cm^{-1}$, $5^2P_{3/2} = 20874.3 cm^{-1}$, $5^2D_{5/2} = 7987.36 cm^{-1}$,  $32^2F_{7/2} = 107.272 cm^{-1}$, and $34^2D_{5/2} = 102.917 cm^{-1}$

The mmWave reflection off the back window of the glass cell creates a standing wave interference pattern, creating a line of fluorescence from the Rydberg atoms with a periodic modulation according to the wavelength of the standing mmWave with electric-field intensity $I_{\text{TRA}} = (1-V cos(2kz + \phi))$. The Rydberg laser is then scanned from -60 MHz to +60 MHz around resonance. For each detuning, a fluorescence image is recorded (typical exposure 0.5 s; vertical binning = 3; no horizontal binning), producing a stack of images across the detuning range.

\begin{figure}[h]
    \centering\includegraphics[width=0.9\textwidth]{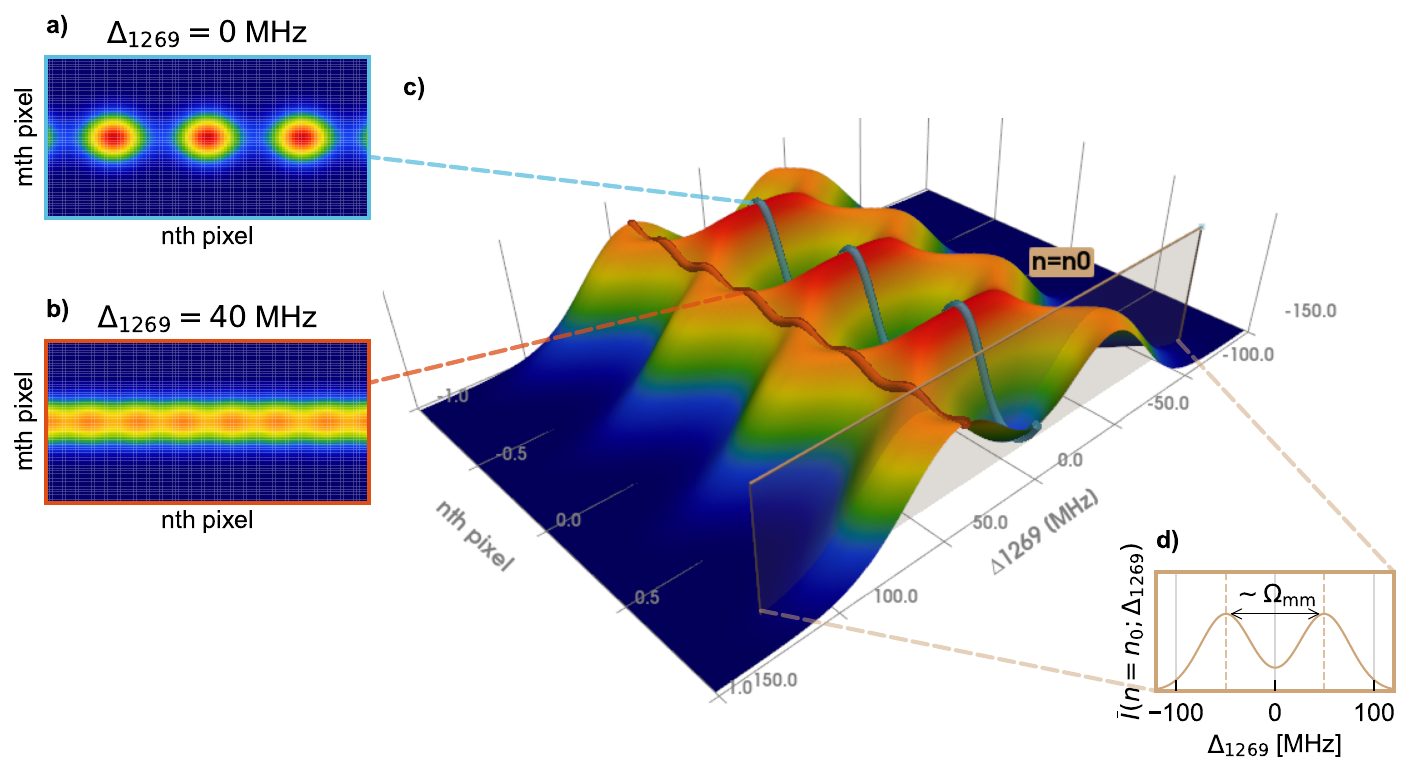}
    \caption{\textbf{a)} Idealized fluorescence photo at Rydberg laser detuning $\Delta_{1269} = 0$ MHz \textbf{b)} Idealized fluorescence photo at Rydberg laser detuning $\Delta_{1269} = 40$ MHz. \textbf{c)} Idealized visualization of fluorescence intensity for 1269 nm laser detuning and position along the laser. Every photo is summed along the $m$th for every detuning, total intensity summed along the z axis. \textbf{d)} An arbitrary position $n_0$ along the laser returns a detuning spectrum analogous to an Autler-Townes splitting in Rydberg-atom receiver schemes with a Rabi frequency $\Rabisig$.}
    \label{visualization}
\end{figure}





\section{Imaging}

The overall electric‑field amplitude profile varies with position in the cell due to a standing wave being formed between the incident mmWave and the reflection from the glass window of the cell. Before every set of measurements, a dark count image is taken. Every image has a pixel count of ($m \times n$). This image consists of the system with all lasers on and locked, with only the mmWave not present in the cell. This is then subtracted from every fluorescence image in the data analysis. Each picture consists of the fluorescence originating from the Rydberg atoms created by the selected detuning of the Rydberg laser, as shown in Fig. \ref{visualization}\textbf{a} and \textbf{b} with the horizontal axis being the $n$th pixels and the vertical axis, the $m$th pixels. We can describe the background-subtracted pixel intensity of the fluorescence at a given detuning in this form: $I_{\text{corrected}}(m,n;\Delta_{1269})$. Since the wavelength of the mmWave, \textasciitilde 2.293~mm, is larger than the beam size of our laser, \SI{700}{\um} beam diameter, we integrate the vertical axis $m$ for each $n$ pixel. We can then recreate an intensity plot, Fig. \ref{visualization}\textbf{c} corresponding to $ \overline{I}(n;\Delta_{1269}) = \sum_{m} I(m,n;\Delta_{1269})$. Now, a slice over the $n$ axis, $I(n=n_0;\Delta_{1269})$, in this new plot gives the intensity only as a function of Rydberg laser detuning, Fig \ref{visualization}\textbf{d}, of which we declare the difference between the two peaks scale as a function of the Rabi frequency of the mmWave wave, similar to an AT splitting measurement.

\section{Data Analysis}

\subsection{Optical correction}
 Because the cell is mounted inside a clear PVC tube, fluorescence emitted away from the cell center experiences refraction and distortion on its way to the camera. This causes a stretching of the image away from the center as well as a slight blurring. To improve spatial calibration, we record an image of a millimeter grid $I_{\text{ref img}}(m,n)$ through the same optics and convert pixel locations to corresponding millimeter distances in our fluorescence composites. Starting with the reference image, the vertical mm separated lines are obtained by first taking the average pixel intensity along the vertical axis $m$. Using the maximum intensity of the new image, it is normalized, resulting in a set of peaks $P = \{n_p\}$. Using the pixel difference between the center-most peak and its nearest neighbor, a uniform grid is created as a function of pixel distance. As each peak position is supposed to be a multiple of 1 mm, a bijective function can be defined to map the pixel position of each peak to a respective mm value. 
 \begin{align*}
     f &: P \rightarrow D \\
     D &= \{x \in \mathbb{Z} | -\frac{\abs{P}}{2} \leq x < \frac{\abs{P}}{2}\} \\
     D &= \{-19,-18,\ldots, 18,19\}\,\mathrm{mm}
 \end{align*}
 We can then plot $\{P,D\}$ and since the peaks pixel position directly corresponds to a millimeter difference, we interpolate using a 3rd order polynomial, $z(n) = an^3 + bn^2 + cn + d$, which becomes the mapping from pixels to $z$ in millimeters. Now $\bar{I}(z;\Delta_{1269})$ can be obtained by applying $z(n)$ to the argument of $\bar{I}(n;\Delta_{1269})$.

\subsection{Fitting}

The readout scheme for Autler-Townes splitting measurements relies on the results of solving the Gorini-Kossakowski-Sudarshan-Linblad (GKSL) master equation from which one can derive that the splitting $s_{A-T}$ of the EIT peak is directly proportional to the Rabi frequency of the incident field ($\Rabisig$).

It is common practice to retrieve the splitting amount from the signal by simplifying the fit to finding the separation of two Gaussian profiles using the least squares method. Such an approach works well for a typical Autler-Townes measurement, where a single probe laser absorption spectrum is analyzed at a time, all fields are in resonance and the field strength is sufficient to separate the peaks, so that for a Gaussian profile of variance $\sigma^2$, the separation $s_{A-T}$ is above $2\sigma$. In the case of fluorescence imaging, there are thousands of spectra to analyze collectively, with the standing wave having local minima below the threshold described above. Furthermore, our measurements were performed in a slightly detuned regime, as can be seen in Fig.~\ref{fig:exp_minmax}, where one of the peaks is slightly stronger.

Under these conditions, retrieving $s_{A-T}$ using Gaussian fits would necessitate more involved techniques, such as forward fits with an additional estimation workflow for recovering the detunings. While this is feasible, the obtained excellent statistical figures of merit do not directly translate to a high confidence of the results being physically accurate, especially as hard to resolve minima are being traversed and it is difficult to verify the spectra individually.

Rather than fine tune the aforementioned workflow, we have determined that the added complexity of solving the master equation is comparable. The model, which is described in greater detail in Appendix A, takes most of the physical parameters into consideration and acts as a coarse self-consistency check as parameters can be compared with the experimental conditions.

The fitting procedure goes as follows: first, we make the claim that fluorescence intensity recorded by the camera scales proportionally with the upper Rydberg population. Under this assumption, we fit the master equation to the plots obtained in Fig. \ref{visualization}\textbf{d}. The Autler–Townes effect is observed as a splitting of the EIT in the fluorescence spectrum of the Rydberg laser. This is proportional to the expectation values of $\bra{4}\rho(\Rabisig,\Delta_{1269})\ket{4}$. In the experiment, the measurement of fluorescence happens some time after the moment the interaction started, which allows the system to relax into a steady state. In the macroscopic picture, the steady state is time independent, $\dot \rho = 0$, which simplifies the calculations significantly. By manually fitting a point $\bar{I}(z_0,\Delta_{1269})$, we obtain the values for all other parameters in the GKSL equation and generate fluorescence spectra by only varying $\Rabisig$, under the assumption that the remainder of the system exhibits negligible or no variations along the cell. These generated spectra can now be directly compared to the spectra found in our experiment over all $z$. Subsequently, we perform least square fits of the profiles, which are now a function of just two parameters, with the sum of squared residuals (SSR) being given by: 
\begin{equation}
SSR(z;A,\Rabisig) = \sum_{\Delta_{1269}} \abs{\bar{I}(z;\Delta_{1269})-A\rho_{44}(\Rabisig;\Delta_{1269})}^2.
\end{equation}
The fit returns estimates of two parameters, $\Rabiest(z)$ and a scaling factor $\hat{A}$. The $\Rabiest(z)$ estimates obtained from Gaussian peak fitting have shown a relative difference of about 15\% in the well-behaved region near the center of the cell, which increased to about 25\% in the local minima. Therefore, further backing for our method was required.


\subsection{Verification}
\begin{figure}[h]
    \centering\includegraphics[width=0.9\textwidth]{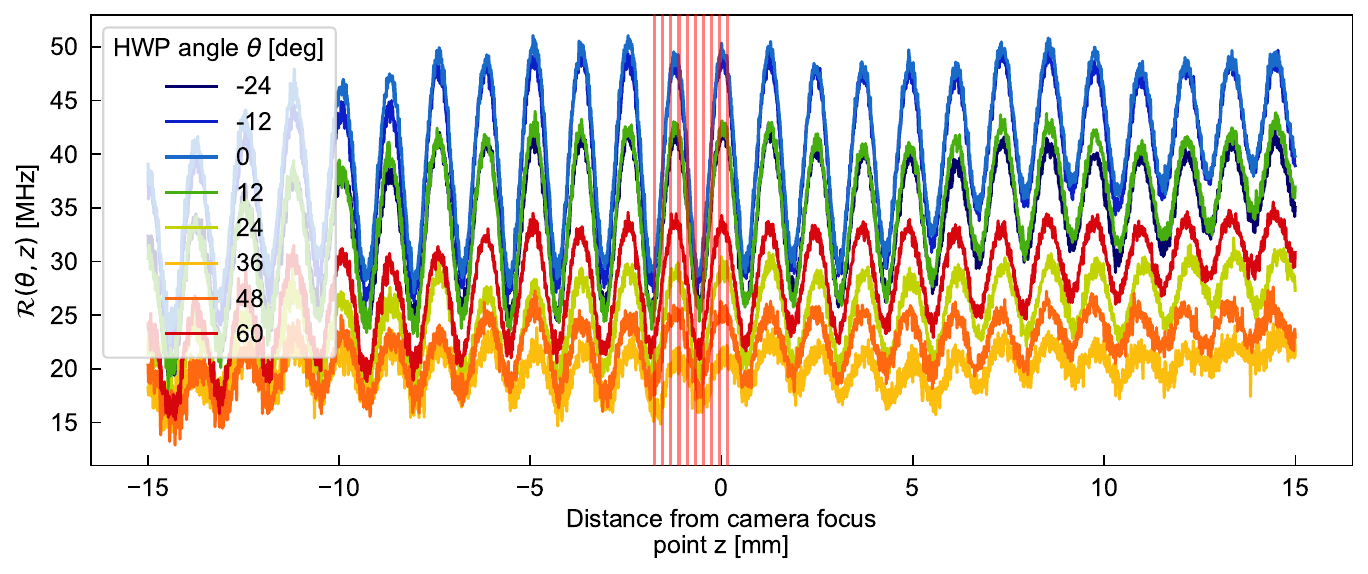}
    \caption{Attenuated Rabi Frequencies $\mathcal{R}(\theta,z)$ are shown with each color indicating a new angle of the HWP $\theta$ corresponding to a change in intensity. The red lines indicate the subset of discrete $z$'s chosen to fit the estimator $\hat{\mathcal{R}}(\theta,z)$ is fitted. These were selected near to the center to decrease image distortion through the glass as well as span a full standing wave.}
    \label{attenuation plots}
\end{figure}

In order to confirm the validity of the $\Rabisig$ fitting, we perform measurements of a well defined attenuation system. Using the polarization control elements in the mmWave path shown in Fig. \ref{fig:levels-setup}\textbf{b}, the signal intensity can be continuously attenuated in a controlled manner. The doublet of a half-waveplate (HWP) and a polarizer (POL) attenuates a linearly polarized wave with a factor:
\begin{equation}
\alpha(\theta) = \cos(4\theta - \theta_0) + \alpha_0,
\label{eq:att}
\end{equation}
where $\theta$ is the HWP rotation angle, $\theta_0$ corresponding to the minimal attenuation angle and $\alpha_0$ modeling an imperfect attenuator with non-zero minimal signal loss. Shown in Fig. \ref{attenuation plots}, multiple image scans were taken with varying $\theta$, written as 
\begin{equation}
    \mathcal{R}(\theta,z) = \alpha(\theta)\Rabisig(z)
\end{equation}

From these scans, a subset of discrete $\Rabisig(z)$ values is taken for each HWP angle $\theta$. This subset of $z$'s is chosen near the center of the cell, as preliminary fits are in the middle of the range of Rabi frequencies, as well as having the least image distortion through the glass. These are then used to estimate the attenuation \textbf{fit} denoted with a hat: $\hat{\mathcal{R}}(\theta,z)$. In this situation, the fit is constrained where the attenuator system, $\hat{\alpha}(\theta)$, must be the same for every $\Rabiest(z)$. 

\begin{figure}[h]
    \centering\includegraphics[width=0.8\textwidth]{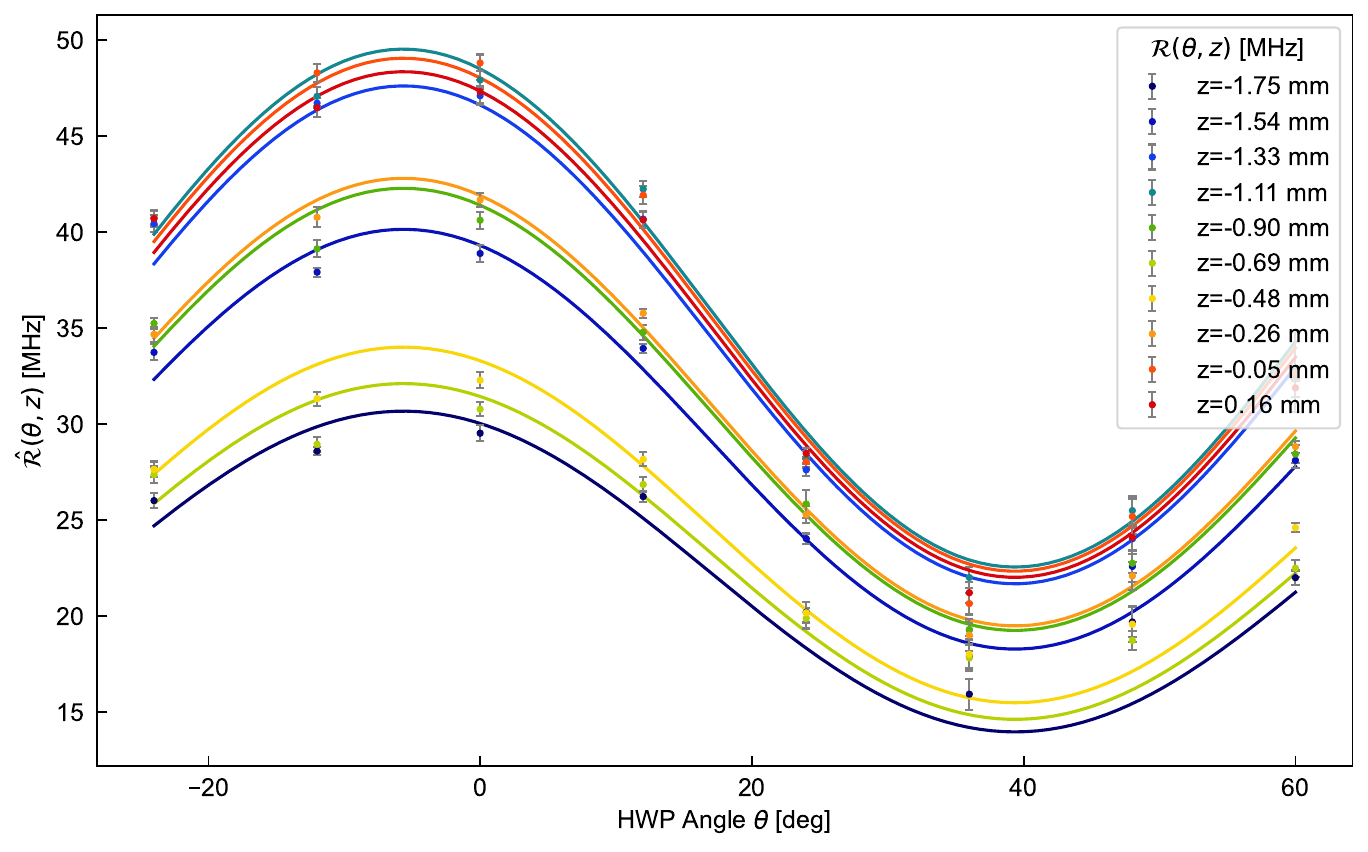}
    \caption{Estimator fit $\hat{\mathcal{R}}(\theta,z)$ using the $\mathcal{R}(\theta,z) = \alpha(\theta)\Rabisig(z)$ obtained at z points from Fig. \ref{attenuation plots}. Shown as different colors, every z position has corresponding measured attenuated Rabi Frequencies $\mathcal{R}(\theta,z)$ at various angles $\theta$. Estimator fit $\hat{\mathcal{R}}(\theta,z)$ is then calculated with the constraint that $\hat{\alpha}(\theta)$, must be the same for every $\Rabiest(z)$ }
    \label{estimator fit}
\end{figure}

In Fig. \ref{estimator fit}, the colored lines represent the estimator fit with the experimental points shown for each z-position at their respective HWP angle $\theta$. To show the validity of the estimator, a parity plot will be introduced comparing the estimator values with the experimental values. 

\begin{figure}[h]
    \centering\includegraphics[width=0.8\textwidth]{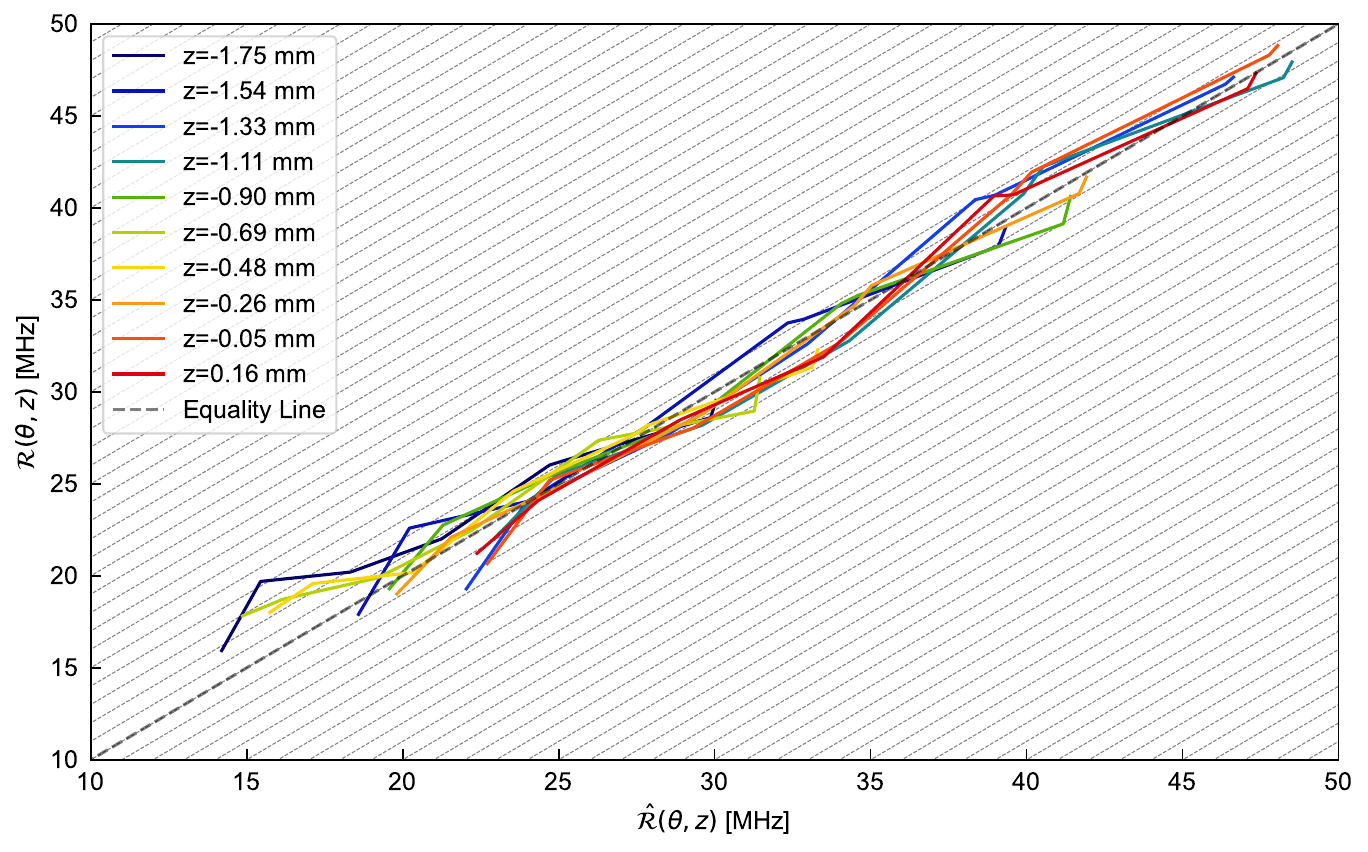}
    \caption{Parity plot of estimator $\hat{\mathcal{R}}$ and experimental values $\mathcal{R}$ where the dashed line represents the ideal case where the estimator reflects the experiment exactly, or in other words, an identity mapping: $\hat{\mathcal{R}}=\mathcal{R}$ }
    \label{parity plot}
\end{figure}

Fig. \ref{parity plot} below uses the estimated attenuated Rabi Frequency $\hat{\mathcal{R}}(\theta, z)$ as an $x$ axis, while the measured attenuated Rabi Frequency points $\mathcal{R}(\theta, z)$ are plotted with their respective values on the $y$ axis. This is analogous to the implementation of an $xy$ mapping for the estimator $\hat{\mathcal{R}}$ and data $\mathcal{R}$. The diagonal line shows the ideal case, an identity mapping $\hat{\mathcal{R}}=\mathcal{R}$, in which the estimator perfectly captures the measured values. This similarity gives confidence that the GKSL fit provides an accurate descriptor of $\Rabisig$.  

\section{Potential applications}
\begin{figure}[h]
    \centering
    \includegraphics[width=\textwidth]{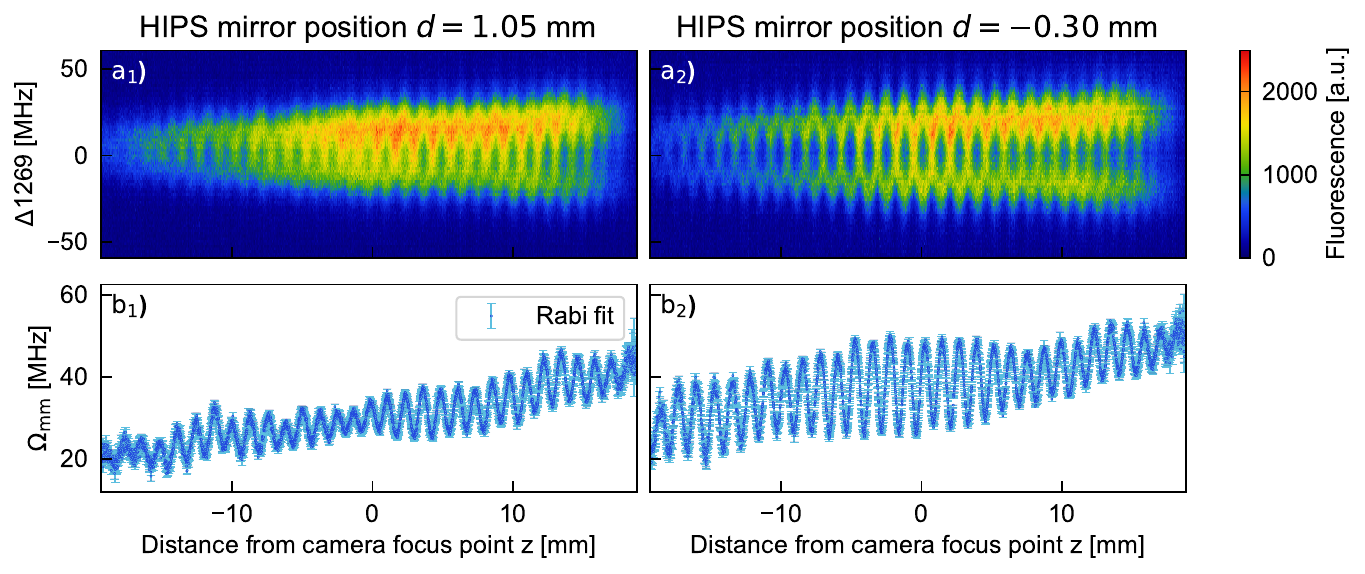}
    \caption{Fluorescence scans with varying HIPS Bragg reflector positions minimizing and maximizing, $\mathbf{a_1}$ and $\mathbf{a_2}$ respectively, standing wave visibility. Fluorescence intensity was normalized for ease of comparison and fitting. The fitted Rabi Frequency via AT-splitting is plotted in $\mathbf{b_1}$and $\mathbf{b_2}$, showing the change in standing-wave amplitude.}
    \label{fig:exp_minmax}
\end{figure}

\begin{figure}[h]
    \centering
    \includegraphics[width=\textwidth]{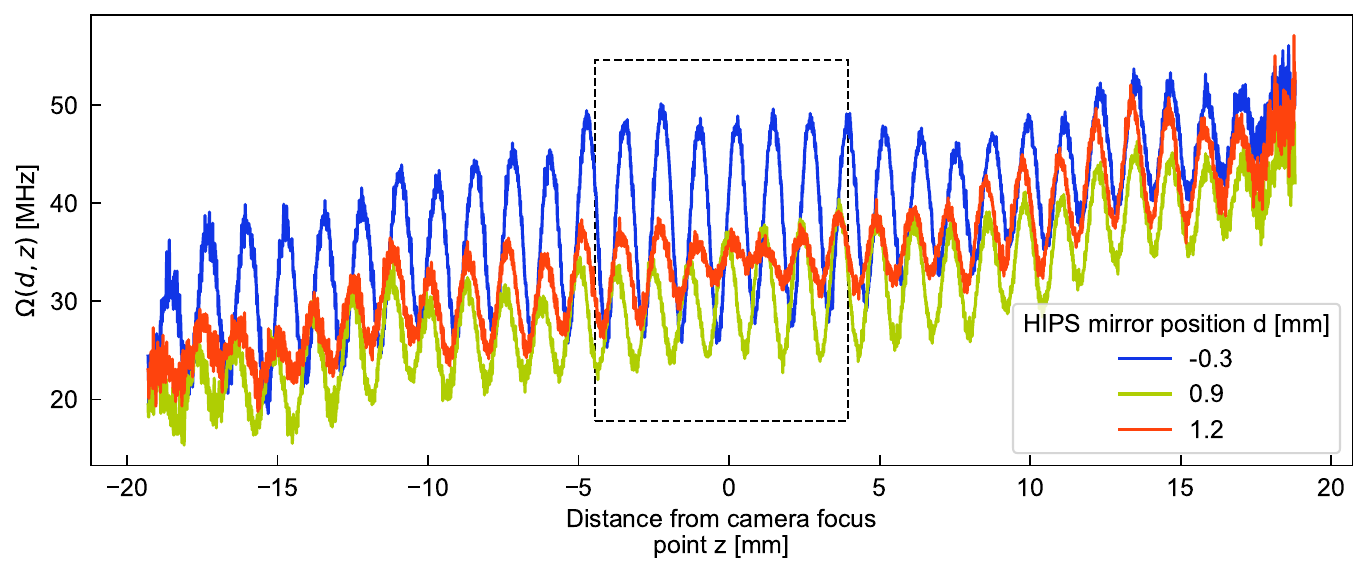}
    \caption{Direct overlay of E-field measurements of various HIPS mirror positions to show local suppression of the field shown in the red box}
    \label{fig:local_suppression}
\end{figure}

As an exploration of mmWave and THz interacting materials, work has been done showing that High Impact Polystyrene (HIPS) has "low, nearly flat absorption across the band..." \cite{freitas_optical_2025} in the high GHz to <1~THz band, as well as being a relatively cheap and 3D printable. While used as a HWP and QWP in the experimental setup, further interest in potential applications led to the development of a simple HIPS Bragg reflector as a method of controlling the back-reflections seen inside the glass cell.
As such, multiple scans were taken to sweep through the standing wave amplitude.  We introduce the HIPS Bragg reflector on the side of the cell opposite the incident mmWave beam. Fig.~\ref{fig:exp_minmax} shows fluorescence composites for reflector positions $d$ that minimize and maximize the fringe visibility, respectively. Adjusting $d$ tunes the reflected amplitude and phase, allowing suppression or enhancement of the standing‑wave contrast inside the cell shown in Fig.~\ref{fig:local_suppression}. While experimental perfect suppression was not achieved, the authors believe that this form of field suppression has potential applications in in-situ electric field measurements. 

\section{Conclusion} 

We have demonstrated fluorescence-based imaging of mmWave electric fields in a warm $\Rb$ vapor cell, with absolute calibration provided by the position-resolved Autler–Townes splitting of a Rydberg EIT resonance. By scanning the Rydberg laser and recording specific fluorescence line images, we reconstruct a background free visualization of the local AT splitting and thus extract the Rabi frequency and mmWave field amplitude along the laser path.

This technique provides a convenient tool for diagnosing and engineering mmWave field distributions in and around vapor cells, including the effects of windows, cavity mirrors, and engineered reflectors. Future work may explore alternative fluorescence channels with shorter lifetimes to enable faster imaging, as well as light‑sheet excitation schemes to obtain two‑dimensional field maps.

\begin{backmatter}

\bmsection{Acknowledgment}
Funded by the European Union (Grant Agreement: 101120422). Views and opinions expressed are, however, those of the author(s) only and do not necessarily reflect those of the European Union or REA. Neither the European Union nor the granting authority can be held responsible for them.
The "Quantum Optical Technologies" (FENG.02.01-IP.05-0017/23) project is carried out within the Measure 2.1 International Research Agendas programme of the Foundation for Polish Science, co-financed by the European Union under the European Funds for Smart Economy 2021-2027 (FENG). This research was funded in whole or in part by the National Science Centre,
Poland grant No. 2021/43/D/ST2/03114.
We thank K. Banaszek for support and discussions.

\bmsection{Disclosures}
The authors declare no conflicts of interest.





\bmsection{Data availability} Data underlying the results presented in this paper are available in  Ref.~\cite{GQCCQT_2026}.

\bmsection{Code availability} Code samples are available in Ref.~\cite{GQCCQT_2026} with more in depth data analysis scripts available from W.~K. upon request.



\section*{Appendix A: Theoretical model}

We follow the usual semi-classical derivations of a single atom influenced by external, classical electric fields. The states do not consider the magnetic sublevels and are purely electronic. While it is known that for strong signal fields the sublevels may distort the signal, that is caused by the presence of mixed polarizations. The state configuration used targets transitions between a minimal and maximal magnetic quantum number, which are outlier transitions. Since our setup ensures the pure sign-matched circular polarization of all fields, the magnetic sublevels may be neglected. Given the state level configuration used, it is a five-level system and after applying the rotating wave and dipolar approximations, we obtain the following Hamiltonian:
\begin{equation}
H = -\tfrac{\hbar}{2}
\begin{pmatrix}
0 & \scriptstyle \Omega^*_{780} & 0 & 0 & 0 \\
\scriptstyle \Omega_{780} & \scriptstyle i(\Gamma_1 +\Gamma_\text{tr}) + 2\Delta_1 & \scriptstyle \Omega^*_{776} & 0 & 0 \\
0 & \scriptstyle \Omega_{776} & \scriptstyle i(\Gamma_2 +\Gamma_\text{tr}) + 2\Delta_2 & \scriptstyle \Omega^*_{1269} & 0 \\
0 & 0 & \scriptstyle \Omega_{1269} & \scriptstyle i(\Gamma_3 +\Gamma_\text{tr}) + 2\Delta_3 & \scriptstyle \Omega^*_{\text{mm}} \\
0 & 0 & 0 & \scriptstyle \Rabisig & \scriptstyle i(\Gamma_4 +\Gamma_\text{tr}) + 2\Delta_4
\end{pmatrix},
\end{equation}
where $\Omega_{\text{wavelength}}$ are Rabi frequencies of respective fields as denoted by the wavelength, $\Gamma_i$ is the decay rate of state $\ket{i}$ and the effect of an atom crossing the interaction region in a finite time is modeled as an additional transit decay $\Gamma_\text{tr}$. Each state experiences an effective detuning corresponding to the total detuning of its driving fields:
\begin{equation}
    \begin{split}
        \Delta_1 &= \Delta_{780}, \\
        \Delta_2 &= \Delta_{780} + \Delta_{776}, \\
        \Delta_3 &= \Delta_{780} + \Delta_{776} + \Delta_{1269}, \\
        \Delta_4 &= 
        \Delta_{780} + \Delta_{776} + \Delta_{1269} + \Delta_{\text{mm}}.
    \end{split}
\end{equation}

The transit time broadening is not the only effect that needs to be considered when working with room temperature vapors. The atoms also experience the Doppler effect, which in the case of our system is the 1D case. This can be modeled by adding extra wave vector based scalar values to the diagonal. The extra term is velocity dependent, yielding:

\begin{equation}
H_v = H + \begin{pmatrix}
0 & 0 & 0 & 0 & 0 \\
0 & \scriptstyle -k_{780} & 0 & 0 & 0 \\
0 & 0 & \scriptstyle -k_{780} + k_{776} & 0 & 0 \\
0 & 0 & 0 & \scriptstyle -k_{780} + k_{776} + k_{1269} & 0  \\
0 & 0 & 0 & 0 & \scriptstyle -k_{780} +  k_{776} + k_{1269} - k_{\text{mm}} 
\end{pmatrix}v.
\end{equation}
where $v$ denotes the velocity class of atoms and the signs of the wave vector magnitudes corresponds to our case of counter propagating probe and signal fields. This Hamiltonian can now be treated with the GKSL (Gorini–Kossakowski–Sudarshan–Lindblad) equation:
\begin{equation}
\begin{split}
   \dot{\rho}_v(t) = &-\frac{i}{\hbar}\left[H_v, \rho_v(t)\right] + \mathcal{L}_\Gamma \left[\rho_v(t) \right], 
\label{eq: gksl}
\end{split}
\end{equation}
to find the density matrix $\rho_v$. The Lindblad superoperator $\mathcal{L}_\Gamma$ introduces repopulation mechanisms to the system. The final solution is obtained by performing Maxwell-Boltzmann weighted statistical averaging:
\begin{equation}
    \rho = \int w(v) \rho_v d v,
\end{equation}

with $w(v)$ being the appropriate weight.

\subsection*{Autler-Townes splitting}
Having obtained the density matrix $\rho$, it is now possible to retrieve the observed Autler-Townes splitting, by extracting the fluorescence spectrum of the Rydberg laser. This is proportional to the expectation values of $\bra{4}\rho(\Delta_{1269})\ket{4}$. In the experiment, the measurement of fluorescence happens after some time from the moment the interaction started, which allows the system to relax into a steady state. The calculations are therefore greatly simplified, since a steady state is time independent, i.e. $\dot{\rho}=0$.


\section*{Appendix B: Cell Holder design}

The cell holder is made of clear PVC with a black paper liner inside to reduce stray light from entering the camera. A flexible Kapton heater is attached to the lower inner part of the PVC pipe with a small hole drilled in the PVC for the heating wire. The cell is then inserted with flexible 3D printed holders made of PETG to resist deformation from heating as well as center the cell. A PVC foam tube is added around the outside of the cell with paper walls, in plane with the cell windows, to prevent stray air currents from cooling the cell windows, with small through holes for the lasers. This holder was created to reduce the amount of metal present around the cell in order to avoid unintentional fields from being scattered and reflected through the cell.

\end{backmatter}


\bibliography{bib}

\end{document}